# Study of Interfacial Tension between an Organic Solvent and Aqueous Electrolyte Solutions Using Electrostatic Dissipative Particle Dynamics Simulations.

E. Mayoral [(1)], E. Nahmad-Achar[(2)]

1 Instituto Nacional de Investigaciones Nucleares, Carretera México-Toluca s/n, La Marquesa Ocoyoacac, Estado de México CP 52750, México. Email address: estela.mayoral@inin.gob.mx
2 Instituto de Ciencias Nucleares, Universidad Nacional Autónoma de México (UNAM), Apartado Postal 70-543, 04510 México D.F. Email address: nahmad@nucleares.unam.mx

## ABSTRACT

The study of the modification of interfacial properties between an organic solvent and aqueous electrolyte solutions is presented by using electrostatic Dissipative Particle Dynamics (DPD) simulations. In this article the parametrization for the DPD repulsive parameters $a_{ij}$ for the electrolyte components is calculated considering the dependence of the Flory-Huggins $\chi$ parameter on the concentration and the kind of electrolyte added, by means of the activity coefficients. In turn, experimental data was used to obtain the activity coefficients of the electrolytes as a function of their concentration in order to estimate the $\chi$ parameters and then the $a_{ij}$ coefficients. We validate this parametrization through the study of the interfacial tension in a mixture of n-dodecane and water, varying the concentration of different inorganic salts ($NaCl$, $KBr$, $Na_2SO_4$ and $UO_2Cl_2$). The case of $HCl$ in the mixture n-dodecane/water was also analyzed and the results presented. Our simulations reproduce the experimental data in good agreement with previous work, showing that the use of activity coefficients to obtain the repulsive DPD parameters $a_{ij}$ as a function of concentration is a good alternative for these kinds of systems.

## I. INTRODUCTION

The change of interfacial tension by the use of additives has an important effect for many industrial applications which involve oil/water interfaces [1]. The performance of this kind of additives is in strong correlation with the characteristics of the medium. In particular the ionic strength, pH, temperature and pressure play an important role in the behavior of these complex systems. Understanding how these conditions modify the interfacial tension is a fundamental task in order to design formulations *ad hoc* for particular systems.

Various experimental data and theoretical models for the calculation of interfacial tension between an organic solvent and aqueous electrolyte solutions have been presented. Aveyard and Saleem [2] reported the surface tensions of different aqueous electrolytes solutions and their interfacial tensions with n-dodecane. They discussed their results in terms of both electrostatic theory and dispersion force theory of interfaces, and presented an expression considering a salt free layer, taking into account the non-ideality of the solution. Desnoyer et al. [3] proposed a model by combining thermodynamical equations with an adsorption model. The interfacial tension was obtained from the isothermal Gibbs equation and the variation of interfacial tension as a function of concentration was modeled using a Langmuir-type adsorption equation. The results obtained present a reasonable correspondence with the experimental data. Li and Lu [4, 5] introduced a prediction model for the calculation of the interfacial tension between an organic solvent and aqueous multielectrolyte solutions. In this model, the activity coefficients for the electrolytes are calculated using the Meissner method [6]. Their results were successfully applied to the prediction of interfacial tensions reported in the literature.

Experimental study of these and more complex systems with electrolytic multi-components is complicated, and requires the design, development and testing of the systems at many different conditions, making this study expensive and limited. To avoid this kind of limitations, numerical simulation offers a viable alternative to help in the study and design of formulations with different components, which could give a better performance even under extreme conditions impossible to handle in the laboratory. Recently, mesoscopic simulation has shown to be an attractive research area for both the academic and industrial communities. Complex industrial formulations have to deal with systems composed of large amounts of various kinds of molecules, interacting at different length and time scales, and in these cases using a *coarse grained* approximation is recommendable. Along these lines, Onuki [7] examined the ion distributions (at low ion densities) around an air-salty water interface in the scheme of the Ginzburg-Landau theory, accounting for electrostatic, solvation effects, and image forces, deriving a number of general functional relations for the dielectric constant of the medium, the image interaction, solvation effects and wetting transitions. In [8] he extended this work for less polar fluids and derived a general expression for the surface tension of electrolyte systems, which contains a negative electrostatic contribution proportional to the square root of the bulk salt density. The excess liquid-liquid interfacial tension between two electrolyte solutions as a function of the ionic strength was studied by Bier et al. [9], finding different regimes dependent on unequal ion partitioning and charge separation. By using a coarse-grained simulation method for complex charged systems, coupling a hydrodynamic description to a free energy functional accounting for the interactions between solvent(s) and charged solutes, Rotenberg et al. [10] studied the transport of charged tracers in charged porous media, the deformation of an oil droplet in water under the effect of an applied electric field, and the distribution of ions at an oil–water interface as a function of their affinity for both solvents.

One of the most important mesoscopic simulation approaches is via dissipative particle dynamics (DPD), introduced some years ago by Hoogerbruge and Koelman [11], who proposed to simulate these complex systems by soft spherical beads interacting through a simple pair-wise potential and thermally equilibrated through hydrodynamics. In this formalism, the beads follow Newton´s equations of motion, involving a repulsive parameter $a_{ij}$ in the conservative part of the force. Therefore, the consistency of the results depends strongly on how the essential molecule's characteristics are included into the interaction parameters $a_{ij}$ between the DPD beads. Later, Groot and Warren established a functional relationship between the conservative repulsion DPD parameters $a_{ij}$ and the Flory-Huggins $\chi$-parameter theory (FH) [12], using the solubility parameters to estimate the cross parameters corresponding with different species. The magnitude of the like-like bead interactions ($a_{ii}$) was related with the isothermal compressibility of the medium at standard conditions. An improved method of parametrization was introduced by Travis et al. [13]. In this, they take advantage of the connection between DPD and the Scatchard-Hildebrand regular solution theory [14, 15], eliminating the limitation of identical repulsive interactions between equal beads, and they link every conservative interaction among beads directly to cohesive energy densities. They presented an alternative method for obtaining the self-interaction parameters and also an alternative to calculating the cross interaction parameter using the cohesive energy densities of the pure fluid.

Electrostatic interactions were introduced into DPD by Groot [16] and by González-Melchor et al. [17]. In both cases the point charge at the center of the DPD particle is replaced by a charge distribution over the particle. Groot solves the problem by calculating the electrostatic field on a grid, and González-Melchor and one of us solve the problem adapting the standard Ewald method to DPD particles. In complex electrostatic systems, the conservative force is composed of the electrostatic contribution apart from the DPD repulsive parameter $a_{ij}$. As is natural to expect, these two contributions are in strong relation with the concentration of charged particles in the system.

A study of the interface between aqueous electrolytes and air has been done by Ghoufi and Malfreyt [18, 19], showing that the surface tension of *NaCl* and *NaF* solutions can be predicted from electrostatic many-body DPD simulations (MDPD). As MDPD is composed of an attractive term and a density dependent repulsive term, the authors developed a multi-scale method to obtain these parameters from atomistic simulations using Flory-Huggins theory.

In this work, we present an alternative methodology to simulate the interfaces between two immiscible liquids (salt solutions and an organic liquid) by using the standard electrostatic DPD simulations, where we only have a repulsive parameter in contrast with the MDPD used by Ghoufi and Malfreyt. To appropriately include the electrostatic information, and in order to obtain realistic parameters, we take into account the concentration of the ions. With this in mind, we introduce a parametrization which considers the dependence of the Flory-Huggins $\chi$ parameters on the electrolyte concentration by using of the Hildebrand –Scatchard regular solution theory via the activity coefficients [14, 15, 20]. The activity coefficients are taken from reported experimental data [21]. The dependence of the $a_{ij}$ parameters with the ion concentration is then considered through the $\chi$ parameter in the DPD simulation. As will be explained later, it arises from the fact that the increase of ions in the system affects the effective interaction in the whole arrangement, and in consequence a variation in the repulsive DPD parameter is expected.

To validate this new parametrization we present a study of the interfacial tension at organic/water interfaces with different concentrations of the inorganic salts $NaCl$, $KBr$, $Na_2SO_4$ and $UO_2Cl_2$, as well as with $HCl$, using electrostatic dissipative particle dynamics simulations. The results are in very good agreement with the experimental data [2].

This paper is divided into six sections. After this introduction (section one), in the second section we describe the fundamental information for the electrostatic DPD simulations using the Ewald method. Section III presents the parametrization of the conservative forces, and we introduce how to obtain these parameters through the activity coefficients as a function of concentration. In section IV the methodology to estimate $\chi$ parameters is given, as well as the details of the DPD simulation. We also show how to calculate the interfacial tension. The results for n-dodecane-water with $NaCl$, $KBr$, $Na_2SO_4$, $UO_2Cl_2$ and $HCl$ systems are presented in section V. Finally, a discussion and the main conclusions are drawn in section VI.

## II. THE DPD ELECTROSTATIC ALGORITHM

Dissipative Particle Dynamics (DPD) is a coarse graining approach which consists in representing a complex molecule such as a polymer or surfactant by soft spherical beads joined by a spring interacting through a simple pair-wise potential and thermally equilibrated through hydrodynamics. The beads follow Newton´s equations of motion:

$$\frac{d\boldsymbol{r}_i}{dt} = \boldsymbol{v}_i; \ \frac{d\boldsymbol{v}_i}{dt} = \boldsymbol{f}_i, \tag{1}$$

where the force $\boldsymbol{f}_i$ on a bead $i$ is constituted by three pair-wise additive components:

$$\boldsymbol{f}i = \sum_j (\boldsymbol{f}_{ij}^C + \boldsymbol{f}_{ij}^D + \boldsymbol{f}_{ij}^R). \tag{2}$$

The sum runs over all nearby particles within a distance $r_c$. The conservative force is defined as

$$\boldsymbol{f}_{ij}^C = \begin{cases} a_{ij}\omega^c(r_{ij})\hat{\mathbf{r}}_{ij} & (r_{ij} < r_c) \\ 0 & (r_{ij} \geq r_c) \end{cases} \tag{3}$$

Here, $a_{ij}$ is a parameter which represents the maximum repulsion between particles $i$ and $j$, $\boldsymbol{r}_{ij} = \boldsymbol{r}_i - \boldsymbol{r}_j$, $r_{ij} = |\boldsymbol{r}_{ij}|$, and $\boldsymbol{r}_{ij} = \boldsymbol{r}_{ij}/r_{ij}$ where $r_i$ denotes the position of particle $i$, and the weight function $\omega^c(r_{ij})$ is given by

$$\omega^c(r_{ij}) = \begin{cases} 1 - \frac{r_{ij}}{r_c} & if \quad r_{ij} < r_c \\ 0 & if \quad r_{ij} \geq r_c \end{cases} \tag{4}$$

This conservative repulsion force derives from a soft interaction potential and there is no hard-core divergence as in the case of the Lennard-Jones potential. This fact makes more efficient the scheme of integration, since such a soft force law allows for a large time step. When we need to introduce a more complex molecule as a polymer, we use beads joined by springs with a spring constant $k$, so we also have an extra spring force given by $\boldsymbol{f}_{ij}^S = -k\boldsymbol{r}_{ij}$ if $i$ is connected with $j$. The dissipative and random standard DPD forces are given by

$$\boldsymbol{f}_{ij}^D = -\gamma\omega^D(r_{ij})(\hat{\boldsymbol{r}}_{ij} \cdot \boldsymbol{v}_{ij})\hat{\boldsymbol{r}}_{ij} \tag{5}$$

$$\boldsymbol{f}_{ij}^R = \sigma\omega^R(r_{ij})(\theta_{ij} 1/\sqrt{\delta_t})\hat{\boldsymbol{r}}_{ij} \tag{6}$$

where $\delta_t$ is the time step of integration , $\boldsymbol{v}_{ij} = \boldsymbol{v}_i - \boldsymbol{v}_j$ is the relative velocity and $\theta_{ij}(t)$ is a random Gaussian number with zero mean and unit variance; $\gamma$ and $\sigma$ are the dissipation and the "noise" strength of a random (thermal) field respectively, $\omega^D(r_{ij})$ and $\omega^R(r_{ij})$ are dimensionless weight functions. These quantities are not independent; they are related by the fluctuation-dissipation theorem as shown in Español and Warren [23]: $\gamma = \frac{\sigma^2}{2k_BT}$ , and $\omega^D(r_{ij}) = \left[\omega^R(r_{ij})\right]^2$, where $k_B$ is Boltzmann's constant and $T$ is the temperature. The units are as follows: $[\gamma] = Kg/s$, $[\theta] = s^{-1/2}$ , $[\sigma] = Kg\ m/s^{3/2}$. As mentioned in the Introduction, we will follow Gonzalez-Melchor et al [17] in solving for the electrostatic interactions by adapting the Ewald sum method. The Ewald sum technique is the most employed route to calculating electrostatic interactions in microscopic molecular simulations [24], but in the implementation to DPD simulations it has the problem that because the electrostatic interactions between DPD particles are soft, the atoms with opposite charge form artificial clusters of ions. González-Melchor et al. solve this problem by combining this standard method with charge distributions over particles. Suppose that we have an electro neutral system constituted by $N$ particles, each one with a point charge $q_i$ and a position $r_i$ in a volume $V=Lx\ Ly\ Lz$. Charges interact according to Coulomb's law and the total electrostatic energy for the periodic system is given by

$$U(r^N) = \frac{1}{4\pi\epsilon_0\epsilon_r}\left[\sum_i \sum_{j>i} \sum_{\boldsymbol{nx}} \sum_{ny} \sum_{nz} \frac{q_i q_j}{\left|\boldsymbol{r_{ij}} + (n_x\boldsymbol{L_x}, n_y\boldsymbol{L_y}, n_z\boldsymbol{L_z})\right|}\right] \tag{7}$$

Where $\boldsymbol{n}=(n_x,n_y,n_z)$, $n_x$, $n_y$, $n_z$ are non-negative integer numbers. $\varepsilon_0$ and $\varepsilon_r$ are the dielectric constants of vacuum and water at room temperature respectively. Then it is possible to decompose the long-range electrostatic interactions in real and reciprocal space getting a short-ranged sum written as

$$U(\boldsymbol{r}^N) = \frac{1}{4\pi\epsilon_0\epsilon_r}\left[\sum_i \sum_{j>i} q_i q_j \frac{erfc(\alpha_\varepsilon r)}{r} + \frac{2\pi}{V}\sum_{\boldsymbol{k}\neq 0}^{\infty} Q(k) S(\boldsymbol{k}) S(-\boldsymbol{k}) - \frac{\alpha_\varepsilon}{\sqrt{\pi}}\sum_i^N q_i^2\right] \tag{8}$$

with

$$Q(k) = \frac{e^{-k^2/4\alpha_\varepsilon^2}}{k^2}, \qquad S(\boldsymbol{k}) = \sum_{i=1}^N q_i e^{i\boldsymbol{k}\cdot\boldsymbol{r}_{ij}}, \qquad \boldsymbol{k} = \frac{2\pi}{L}(m_x, m_y, m_z) \tag{9}$$

$\alpha_\varepsilon$ is the parameter that controls the contribution of the real space, $k$ is the magnitude of the reciprocal vector **k**, $m_x, m_y, m_z$ are integer numbers, and $erfc(\alpha_\varepsilon r)$ is the complementary error function. It is important to notice that in standard DPD methodology the conservative force is mathematically well defined at $r = 0$, allowing for a full overlap between particles, but the electrostatic contribution diverges at $r = 0$ leading to the formation of ionic pairs; that is the reason why a charge distribution is used in our methodology [17].

## III. PARAMETRIZATION OF THE CONSERVATIVE FORCES

From equations 3, 5 and 6, we can see that in the DPD simulation we have several parameters to establish: $a_{ij}$, $\sigma$ and $\gamma$. The most important one is the conservative force parameter $a_{ij}$, because it contains all the chemical information for each component in the system; the noise and dissipative parameters ($\sigma$ and $\gamma$) correspond to the temperature and fluid viscosity. According with Groot and Warren [12], in a mono-component system there is a simple relationship between the conservative force parameter for equal species $a_{AA} = a$, and the inverse isothermal compressibility. The compressibility of the system is defined by $\kappa^{-1} = 1/nk_BT\kappa_T = 1/k_BT(\partial p/\partial n)_T$ where $n$ is the number density of molecules and $\kappa_T = (\partial p/\partial n)_T$ is the usual isothermal compressibility. The pressure $p$ in the system could be obtained using the viral theorem and for this case is $p = \rho k_BT + \alpha a\rho^2$, where $\rho$ is the density, $\alpha = 0.101$ for $\rho > 2$ and $a$ the DPD repulsive parameter. Then we have $\kappa^{-1} = 1 + 2\alpha a\rho/k_BT \approx 1 + 0.2a\rho/k_BT$. Following [25] we take into account the degree of coarse graining characterized by the number $N_m$ of molecules contained in a DPD particle, obtaining:

$$a = k_BT(\kappa^{-1}N_m-1)/2\alpha\rho_{DPD}, \tag{10}$$

where $\rho_{DPD}$ is the DPD density for the system and is usually set to as $\rho_{DPD} = 3$. In this mono-component DPD case, the virial free energy density $f_v$ is given by $f_v / k_BT = \rho \ln\rho - \rho + \alpha a\rho^2/k_BT$.

For a mixture of two components A and B, assuming $\alpha$ independent of $\rho$ and $a$, the virial pressure $p$ [26] is given by:

$$p = \frac{\alpha k_BT\rho^2}{R_c^3}\left[a_{AA}\phi^2 + 2a_{AB}\phi(1-\phi) + a_{BB}(1-\phi)^2\right], \tag{11}$$

where $\phi$ is the volume fraction of component A and $(1 - \phi)$ the volume fraction of component B. The corresponding virial free energy density for this system is

$$\frac{f_v}{\rho k_BT} = \frac{\phi}{N_A}\ln\phi + \frac{(1-\phi)}{N_B}\ln(1-\phi) + \frac{\alpha(2a_{AB} - a_{AA} - a_{BB})\rho}{k_BT}(\phi)(1-\phi) + cte, \tag{12}$$

with $\rho = \rho_A + \rho_B$ and $a_{AB} = a_{BA}$.

In order to establish a relationship between $a_{ij}$ and the physicochemical characteristics of a real system, the Flory-Huggins (FH) theory is commonly used. This is a simple lattice model based on occupations of a lattice where we have both, solvent and polymer molecules placed on each site [27, 28]. One has exclusively and uniquely a polymer segment or a solvent molecule per lattice site. The standard FH mean field approximation substitutes the single occupancy constraint by site occupancy probability arguments, which give up a mean field free energy of mixing constituted by a combinatorial entropy and a mean field energy of mixing : $\Delta F_{MIX}^{MF} = \Delta S_{MIX}^{MF} + \Delta H_{MIX}^{MF}$. With this, the free energy per unit volume for a mixture of two polymers A and B could be written as:

$$\frac{\Delta F_{MIX}^{MF}}{Nk_BT} = \frac{\phi}{N_A}\ln(\phi) + \frac{(1-\phi)}{N_B}\ln(1-\phi) + \chi(\phi)(1-\phi) \tag{13}$$

where $\phi$ and $1-\phi$ are the volumetric fractions for the $A$ and $B$ components respectively, $N_A$ and $N_B$ are the number of monomers $A$ and $B$ respectively and $N = N_A + N_B$ the total amount of monomers in the system. In this equation, the first two terms of the right hand side contain the information of the energy of the pure components and correspond with the entropic contribution $\Delta S_{MIX}^{MF}$. Therefore, the entropy of mixing does not arise from coulombic interactions, the first and second terms in the right hand side of equation 13 are appropriate for the description of electrolytic systems also [20]. The third one involves the excess energy produced by the mixture (that is $\Delta H_{MIX}^{MF}$); the $\chi$- parameter then tells us how alike the two phases are, and is known as the Flory-Huggins interaction parameter. In the FH mean field theory this parameter $\chi$ is written in terms of the nearest-neighbor interaction energies $\epsilon_{ij}$ as $\chi_{12} = z(\epsilon_{11} + \epsilon_{22} - \epsilon_{12})/2k_BT$, where $z$ is the lattice coordination number. Since the FH model admits only the presence short-range forces, discrepancies arise when using the model in presence of long-range forces. Corrections have been introduced by considering an ionization equilibrium between counter-ions and electrolyte (even for multivalent ions) and the electrostatic excluded volume between the electrolytes modeled at the level of the Debye-Huckel approximation [29, 30, 31]. In practice $\chi_{12}$ is considered as a phenomenological parameter, and the free energy of mixing becomes independent of any parameters proper to the lattice model. As an alternative, an adequate estimation of this quantity could be done by using the Hildebrand-Scatchard regular solution theory [14, 15, 32], which is defined as one for which the entropy of mixing is given by an ideal expression but the enthalpy of mixing is non-zero and is the next simplest approximation to the ideal solution. This approach constitutes an appropriate way to consider the coulombic contribution in the enthalpy of mixing via the activity coefficients [20] in electrolyte solutions as we will show in the next section.

Comparing equations (12) and (13) Groot and Warren [12] proposed that the repulsive parameters $a_{AB}$ in the DPD simulation can be obtained using the χ-Flory-Huggins parameter as

$$\chi_{AB} = \frac{\alpha(2a_{AB} - a_{AA} - a_{BB})\rho}{k_B T}. \tag{14}$$

**III.I Interaction parameters as functions of the salt concentration by using the activity coefficients**

Although the original formulation of FH mean field theory gives the interaction parameter $\chi_{12}$ as independent of the concentration, proportional to $T^{1}$ and energetic in origin, comparisons with experiments show that phenomenological $\chi_{12}$ depends on polymer concentrations ($\xi$) and contains both energetic and entropic contributions $\chi$ ($T$, $\xi$) [29, 31, 33, 34]. Some corrections to the FH mean field theory have been presented, as mentioned above, and a more complex effective Flory-Huggins parameter $\chi_{eff}$ can be estimated via the Hildebrand-Scatchard regular solution theory [39]. Muthukumar et al.[22] have argued that the assumption of a constant polymer charge throughout the phase diagram is erroneous, recognizing that it can vary when polymer concentration, temperature, and other experimental parameters are changed. This suggests that a correct parametrization in our electrostatic DPD system must therefore take into account the dependence of the repulsive parameters for the solvated ions $a_{ij}$ with the salt concentration $\xi$. That this should be so may be understood as follows: when we perform a coarse graining, the volume of a DPD particle does not usually encompass a full molecule or polymer; thus, for instance, although for dodecane our DPD particle contains only a butane fragment, one does not construct dodecane from the union of butane particles, and the interaction between the DPD dodecane particles and water does not correspond with the χ parameter of butane with water; the χ parameter employed to estimate the DPD repulsive parameter $a_{ij}$ should be that of the full dodecane molecule because its behavior is that of the global joined units which affect the electronic distribution throughout. In this case, the "monomeric" units which constituted the dodecane "polymeric" molecule interact through short-range (covalent bond) forces.

When considering solvated $Na^{+}$ or $Cl^{-}$ ions, their concentration is given precisely by the amount of solvated ionic particles present, which corresponds effectively with the amount of "monomeric" solvated ionic units. These are in effect the individual DPD units, which in this case are not covalently joined but are subject to long-range electrostatic forces. Then, the presence and quantity of "monomeric" solvated ions affect the global properties of the network and their corresponding $\chi$ parameter should take into account the whole electrolytic entity (i.e., the total number of ions present); as a consequence the $a_{ij}$ parameters in DPD must be considered as corresponding with the whole electrolytic system, thus forcing a dependence on ion concentration (number of units). This dependence can be obtained directly through the $\chi(T, \xi)$ parameter via the activity coefficients [20] as will be shown below.

Our model includes DPD particles for each species: water, solvated $Na^+$ ions, solvated $Cl^-$ ions, and dodecane segments. The free energy dependence on concentration emerges naturally from the varying number of DPD beads of each type and also from the interactions between DPD beads obtained from the $\chi$ parameter. In fact, in the case of solvated ion particles, the effective $\chi$ parameter in eq.(13) takes into account the number of ions and their long-range interactions. For dodecane-water, the effective $\chi$ parameter takes into account the connectivity constraint in dodecane and the number of "butane monomers" constituting it. The Hookean force between DPD beads, in this latter case, takes care of the short-range covalent links between butanes, to form dodecane. The form of the interaction parameters $a_{ij}$ take into account the electronic distribution in a dodecane molecule, which differs from that of butane, and also take into account the effective solvated ionic array which differs from that of a single ion.

As we have mentioned in the Introduction, in the electrostatic DPD approximation not only the charges produce important modifications to the conservative force but also the repulsion parameters do. As a consequence, if a constant force parameter is used in the simulation trying to justify the increment in interfacial tension only from the ionic charge present, one will not reproduce in an adequate way the experimental results, especially for high concentrations where the effect of the charge is more important. Similar results have been found in the case of the surface tension studies by using MDPD developed by Ghoufi et al. [16]. Furthermore, one could not explain the abatement of interfacial tension produced by $HCl$ (see results below) even though its ionic charge is the same as that for $NaCl$.

The chemical potential $\mu_w$ for the water ($w$) component may be obtained by differentiating the free energy per unit volume, for the mixture $w + e$ , with respect to the number $N_w$ of water molecules, one gets

$$\frac{\mu_w}{k_B T} = \ln(\phi) + \chi(1-\phi)^2 \tag{15}$$

Similarly, for $\mu_e$ we have

$$\frac{\mu_e}{k_B T} = \ln(1-\phi) + \chi(\phi)^2 \tag{16}$$

where $\phi$ and $1- \phi$ are the volumetric fractions for the $w$ (solvent) and $e$ (electrolyte) components respectively. The activity for the electrolyte $e$, $\alpha_e$, is given by

$$\ln(\alpha_e) = \frac{\mu_e - \mu_e^\theta}{RT} \tag{17}$$

where $\mu_e^\theta$ denotes an arbitrarily chosen zero for the component $e$ and is called the standard chemical potential of $e$. The electrolyte activity can be written as $\alpha_e = (x)^x (y)^y (\alpha_e^0 m)^z$. Here $x$ and $y$ are the stoichiometric coefficients of the cation and the anion, and $z = x + y$. $\alpha_e^0$ is the mean activity coefficient of the electrolyte and $m$ is the molality of the electrolyte. Then the $\chi$ parameter for the solvent and the electrolyte could be obtained by

$$\chi = \frac{\ln \alpha_e - \ln(1-\phi)}{\phi^2}. \tag{18}$$

Notice that $\chi$ is a concentration dependent parameter, parametrized by the activity of $e$ at different concentrations. With this expression the scaling of $\chi$ with the quantity of ions present can be studied; the results can be observed in fig 1(a). The behavior of this quantity as a function of the concentration $\xi$ follows a power law ( $\chi \sim \xi^{\tau}$) with characteristic scaling exponents $\tau$ depending on the kind of salt (fig 1(b)). There are different alternatives to obtain the activity coefficients of strong electrolytes in aqueous solution [5, 6]. Here we have used the reported experimental values [21] for *NaCl, KBr,* $Na_2SO_4$ and $UO_2Cl_2$.

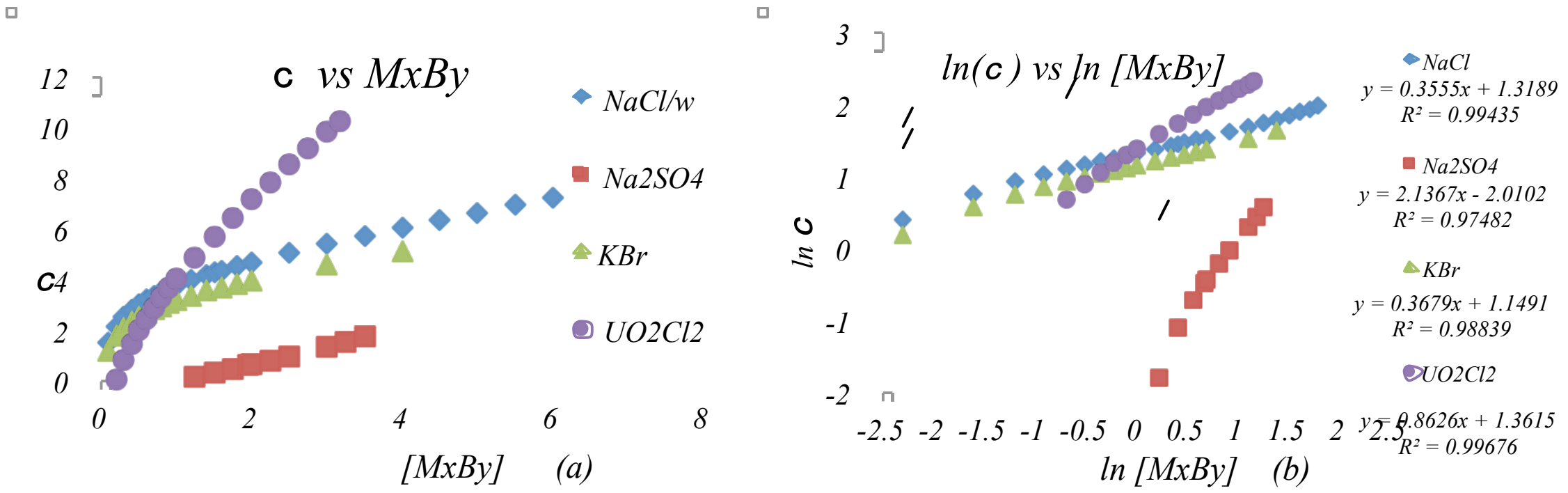

**Figure 1.** Scaling of $\chi$ with the concentration $\xi$ of different inorganic salts in water: *NaCl,* $Na_2SO_4$, *KBr* and $UO_2Cl_2$. (Experimental values for the activity at different concentrations taken from reference [21].)

The repulsive DPD parameter $a_{ij}$ could be obtained using equations 14 and 18, $a_{AB}$ may be obtained from

$$(\delta_A - \delta_B)^2 = -r_c^4 \alpha [\rho_A^2 a_{AA} + \rho_B^2 a_{BB} + 2\rho_A \rho_B a_{AB}] \tag{19}$$

whenever $a_{AA}$ and $a_{BB}$ are known [26]. These may be taken to be equal following the Groot and Warren [12] approximation or may be determined by relating the Hildebrand solubility parameters to the heat of vaporization, following Travis et al. [13]. However, ionic species which have a very large heat of vaporization, such as *NaCl*, lead to large repulsive interaction parameters not suitable for DPD simulations. We therefore take like-like interaction parameters to be equal: $a_{AA} = a_{BB}$. Figure 2 shows these values.

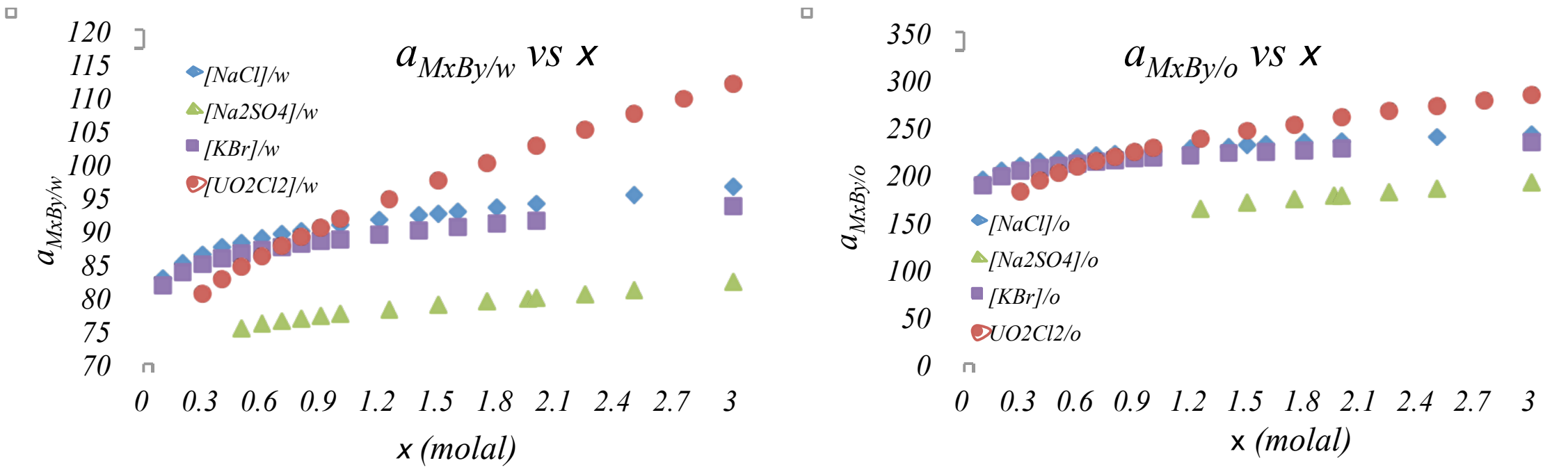

**Figure 2.** $a_{MxBy/w}(\xi)$ and $a_{MxBy/o}(\xi)$

As we can see, the dependence of $a_{ij}$ with the concentration of electrolyte obtained presents a significant difference between the $a_{ij}$ values at different concentrations ($\Delta a_{ij} \approx 1$). In the case of monomeric mixtures the surface tension obtained by DPD simulations is very sensitive to this variation [35], we therefore expect an important deviation in the interfacial tension for these cases.

## IV. DPD SIMULATIONS OF ORGANIC/AQUEOUS ELECTROLYTE SOLUTIONS.

### IV.I Obtention of Flory-Huggins $\chi$ Parameters for the System.

The system that we want to study is a mixture composed by water ($w$), an organic solvent ($o$) and an inorganic salt or electrolyte ( $e = M_xB_y$) at different concentrations. We have considered that the salt $M_xB_y$ is completely dissociated, that is

$$M_xB_y + zH_2O \rightarrow xM^{y+} + yB^{x-} + zH_2O, \tag{20}$$

where $x$ and $y$ represent the stoichiometric coefficients for the cation ($M^{y+}$) and for the anion ($B^{x-}$) respectively and $z = x + y$. As there is no association between the components, we consider that the species interact with each other by pairs $w/o$, $w/M^{y+}$, $w/B^{x-}$, $o/M^{y+}$, $o/B^{x-}$ and $M^{y+}/B^{x-}$ [42]. Additionally we assume that the salt is presented only in the aqueous phase. This could be explained according to the model proposed by Aveyard et al. [2], since the salts that we consider here are negatively adsorbed [4, 5] at the liquid-liquid interfaces; then we can model our system assuming that the concentrations of oil and water quickly drop down to zero at the interface. With this in mind, we can consider three phases: a) A water rich phase, composed of water ($w$), solvated cation ($M^{y+}$) and solvated anion ($B^{x-}$); b) a salt-free interface between the oil and the aqueous electrolyte solution; and c) a segregated phase composed only by oil. According with the traditional theory, the Flory-Huggins interaction parameter between A and B, $\chi_{AB}$, is related with the heat of mixing of the component (A) with the component (B) [27, 28]. If we assume that the heat of mixing in this case is given by the Hildebrand-Scatchard [14, 15, 32] regular solution theory, the $\chi_{AB}$ parameter could be obtained using the solubility parameters ($\delta_A$ and $\delta_B$) for the pure components in the mixture, by the relation

$$\chi_{AB} = \frac{v_{AB}}{RT}(\delta_A - \delta_B)^2. \tag{21}$$

Here $v_{AB}$ is the partial molar volume. It is important to notice that this approximation is valid for non-polar components, but it has been used in polar systems with reasonable success [36]. The $\delta$ solubility parameter could be estimated by using its definition which, according to Hildebrand [32], is the square root of the cohesive energy density, which in turn is defined as the energy of complete vaporization divided by the condensed molar volume. With these considerations and using equation (21) we can express the $\chi$ parameters by pairs in our multicomponent system as

$$\chi_{ew} = \frac{v_{ew}}{RT}(\delta_e - \delta_w)^2 \tag{22}$$

$$\chi_{wo} = \frac{v_{wo}}{RT}(\delta_w - \delta_o)^2 \tag{23}$$

$$\chi_{eo} = \frac{v_{eo}}{RT}(\delta_e - \delta_o)^2. \tag{24}$$

The subscripts *w, e* and *o* correspond to water, electrolyte (cation or anion), and organic interaction respectively. We can express $\chi_{eo}$ as a function of the other two parameters $\chi_{ew}$ and $\chi_{wo}$. Taking the square root of equations (22) and (23) and considering $v_{ew} = v_{wo} = v_{eo} = v_m$ [37, 38] we have:

$$\pm\sqrt{\frac{RT\chi_{ew}}{v_m}} = \delta_e - \delta_w. \tag{25}$$

$$\pm\sqrt{\frac{RT\chi_{wo}}{v_m}} = \delta_w - \delta_o. \tag{26}$$

Adding both equations we obtain

$$\left[\sqrt{\chi_{ew}} + \sqrt{\chi_{wo}}\right]^2 = \frac{v_m}{RT}(\delta_e - \delta_o)^2 \equiv \chi_{eo} \tag{27}$$

This expression allows us to estimate the $\chi_{eo}$ parameter.

The solubility parameters $\delta$ for the water and n-dodecane were calculated performing atomistic dynamics simulations in order to obtain the cohesive energy density $E_{coh}$ and then the $\delta$ values. To do this, periodic cells of amorphous fluids structures were constructed using the Amorphous Cell program of Materials Studio. The dimension of the box was in both cases of *25 Å* on each side. *NpT* dynamics was performed first to equilibrate the density in the system. The COMPASS force field was used in order to describe the interatomic interactions. The Discover Molecular Dynamics engine was used to evolve the systems, generating structures statically independent. The results were compared with the experimental data obtaining a very good agreement with the reported values [39]. The cross parameters $a_{WD}$ between water and n-dodecane was calculated by means of the Flory-Huggins $\chi$ parameters obtained by the solubility parameters $\delta_W$ and $\delta_D$ for water and n-dodecane respectively using the relationship given by equation 21. Where $v_{DPD}$ is the volume of a DPD bead $v_{DPD} = 3\ v_W = 90\ Å^3$ and with $\chi = 0.286(a_{WD} - a)$ then $a_{WD} = 151.537$.

| Molecule | $V_{molecule}$ Å$^3$ | Mean Density (gr/cm$^3$) | Cohesive Energy *25°C* (J/m$^3$) | Solubility parameter δ 25°C ((J/cm$^3$)$^{0.5}$) (Calculated) | Solubility parameter δ 25°C ((J/cm$^3$)$^{0.5}$) (Exp) |
|---|---|---|---|---|---|
| n-dodecane (D) | *319.64* | *0.74* | *2.65E+08* | *16.26* | *16.2* |
| Water (W) | *30.81* | *0.97* | *2.20E+09* | *46.95* | *47.9* |

**Table 1 .**Solubility parameter and cohesive density energy by atomistic simulation at $T=25$ °C (298 K) and $P=1$ atm. The experimental results were taken from ref [39].

## IV.II DPD Simulation Details

All simulations were carried out using a DPD electrostatic code following Gonzalez-Melchor et al. [17]. Dimensionless units (denoted with an asterisk) were used, the dimensionless number density was obtained as $\rho^* = \rho r_c^3$ and the dimensionless repulsive parameters as $a^*_{ij} = a_{ij} r_c / k_B T$. The masses were all equal to 1 and the total average density in the system was $\rho^* = 3.0$ in order to have a quadratic equation of state independent of the magnitude of the conservative force parameter [12]. Higher values of the density than *3* will raise the computing cost significantly. The constants in the dissipative and random forces in equation (1) were set to $\gamma = 4.5$ and $\sigma = 3$ in order to keep the temperature fixed at $k_B T = 1$. A reduced time step of $\Delta t^* = \Delta t(k_B T/m r_c^2)^{1/2} = 0.04$ was used during the simulation. The real forces in the Ewald sums were truncated at $r_c^{real} = 3.0 r_c$ with $\alpha_\varepsilon = 0.15$ Å$^{-1}$. For the reciprocal part, the sum with a maximum vector $\mathbf{k_{max}}$ = (5, 5, 5) was considered and the value of $\beta^* = r_c/\lambda = 0.929$ was established (where $\lambda$ is the decay length of the charge present in the Slater type charge density). In each simulation *25* blocks with *10 000* steps were performed and the properties of interest were calculated by averaging over the last 10 blocks. The total number of DPD atoms in all simulations were *4 500* and the box dimensions were $Lx = 15$, $Ly = 10$, $Lz = 10$. Periodic boundary conditions were imposed in all directions.

We studied the system constituted by a 50:50 mixture of DPD particles of water and n-dodecane, with salt added $M_xB_y$ ($NaCl$, $KBr$, $Na_2SO_4$, $UO_2Cl_2$) or with $HCl$. Three water molecules were represented by one DPD bead, corresponding to a coarse graining of $N_m = 3$. The n-dodecane molecule was represented by *3* identical DPD beads joined by Hookean springs with a harmonic spring constant $k=100$ and an equilibrium distance of $r_e = 0.7$ Å. Solvated ions ($M^{y+}$ or $B^{x-}$) were also represented as one separated DPD bead with its respectively charge distributed as in reference [16, 17]. The concentrations of these salts were varied.

## IV.III Calculation of Interfacial Tension

A property of interest for the validation of the simulation results obtained by the use of the parametrization proposed in this work is the interfacial tension, which can be compared directly with the experimental data. In the case of planar interface the surface tension is given by

$$\gamma = \int_{-\infty}^{\infty} dz [P_{zz}(z) - P_T(z)], \tag{28}$$

where $P_{zz}(z) = P$ and $P_T(z)$ are the normal and tangential components of the pressure, and $P$ is the equilibrium pressure. The use of this expression involves the computation of the components of the pressure tensor as a function of the distance from the interface. This quantity is obtained by means of the virial route [17, 40, 41] as shown in the next equation:

$$P_{zz} = \sum_{i=1}^{N} m_i \boldsymbol{v}_i \ \boldsymbol{v}_i + \sum_{i=1}^{N} \sum_{j>i} z_{ij} \, F_{ij,z}^{C}, \qquad (29)$$

where the first and second terms represent the kinetic and (conservative) interaction contributions, respectively; $m_i$ represents the mass of each DPD bead, $z_{ij} = z_i - z_j$ is the coordinate difference of particles $i$ and $j$ in the $z$-direction, and $F_{ij,z}{}^{C}$ is the $z$-component of the total conservative force between particles $i$ and $j$. Equivalent expressions may be used for $P_{xx}$ and $P_{yy}$, simply by replacing $z$ for $x$ and $y$ respectively.

## V. RESULTS

### V.I DPD Simulation for n-dodecane/water with *NaCl, KBr,* $Na_2SO_4$ and $UO_2Cl_2$

Using the parameters shown in figure 2, for water, n-dodecane and the electrolyte (*NaCl, KBr,* $Na_2SO_4$ and $UO_2Cl_2$) we performed DPD electrostatic simulations for the system n-dodecane/water with salt added at different concentrations using the DPD electrostatic methodology. The result for the interfacial tension between n-dodecane and water without any salt in DPD units was $\gamma_{DPD} = 6.7 \pm 0.0013$. The results for the change in the interfacial tension between n-dodecane and water when an electrolyte is added, is shown in figure 3 for each electrolyte. As expected in all cases, an increase in the interfacial tension between n-dodecane and water is obtained when we add an inorganic salt.

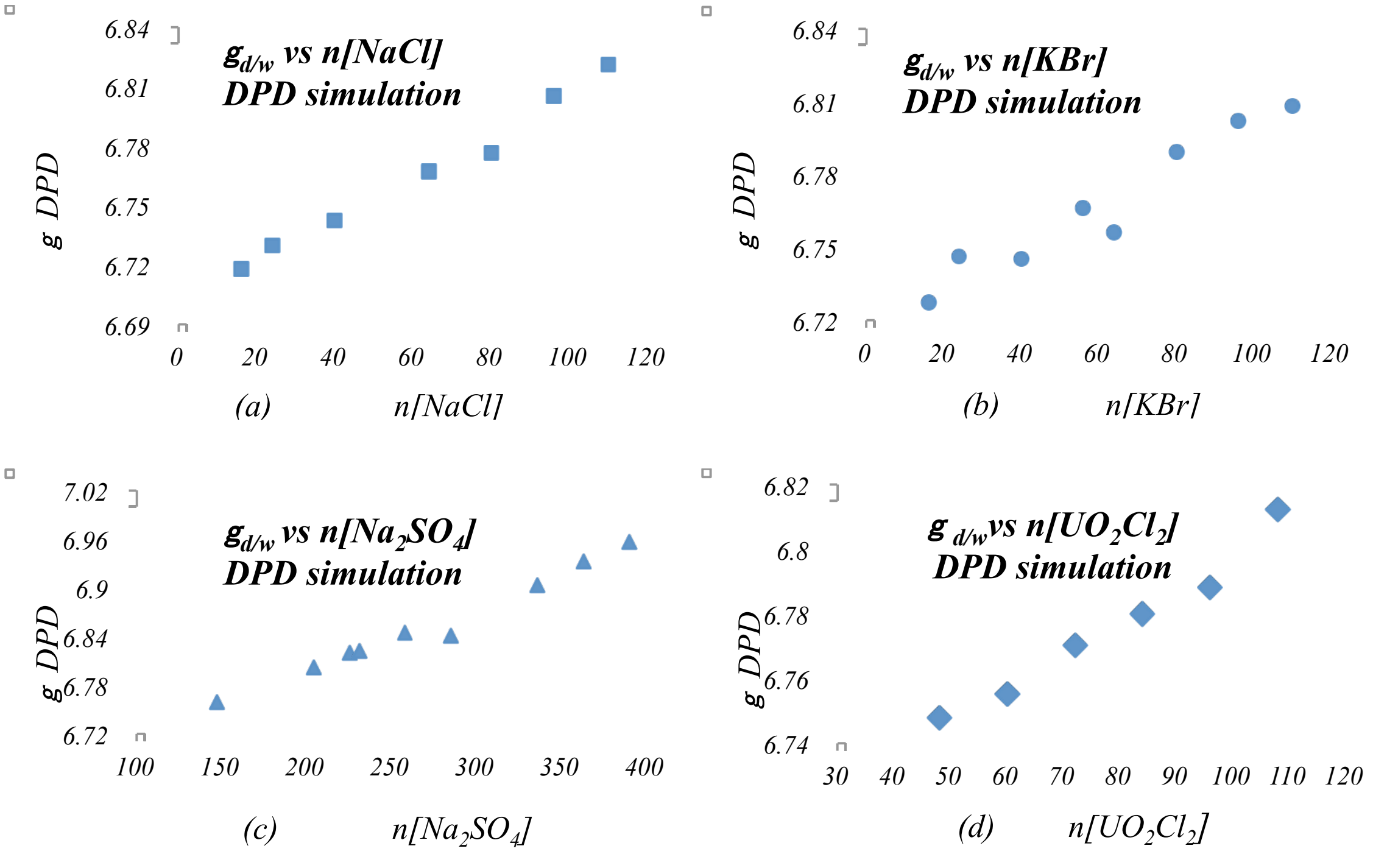

**Figure 3.** Interfacial Tension between n-dodecane and water with (a) *NaCl,* (b) *KBr,* (c) $Na_2SO_4$ ,(d) $UO_2Cl_2$, added, obtained by DPD electrostatic simulations. $\gamma_{d/w}$ is the interfacial tension in DPD units and $n[M_xB_y]$ is the number of DPD ions added.

Considering a total dissociation according with equation (20), the correspondence between the real concentration in molal units [mol $M_xB_y$/ Kg $H_2O$] and the DPD concentration of ions added could be calculated as:

$$m^* = \left[\frac{n\ molec\ M_xB_y}{z\ n\ molec\ H_2O}\right]\left[\frac{1000\ gH_2O}{Kg\ H_2O}\right]\left[\frac{mol\ H_2O}{18\ g\ H_2O}\right]\left[\frac{N_A\ molec\ H_2O}{mol\ H_2O}\right]\left[\frac{mol\ \ M_xB_y}{N_A\ molec\ M_xB_y}\right]$$

$$= \left[\frac{n\ M_xB_y}{z\ n\ H_2O}\right]\left[\frac{1000}{18}\right]\left[\frac{mol\ \ M_xB_y}{Kg\ H_2O}\right] \tag{30}$$

where $N_A$ the Avogadro's number. Using the correspondence between real units given by $\gamma_{calc} = (\ \kappa T\ /\ r_c)\gamma_{DPD}$ we can obtain the increase of interfacial tension between n-dodecane and water, $\Delta\gamma_{dw} = \gamma^0 - \gamma_{calc}$ in [dyn/cm] with $\gamma^o$ the interfacial tension between water and n-dodecane without salt. This result can be compared directly with the experimental data as is shown in figure 4. We can observe a very good agreement between our results and the experimental data reported [2].

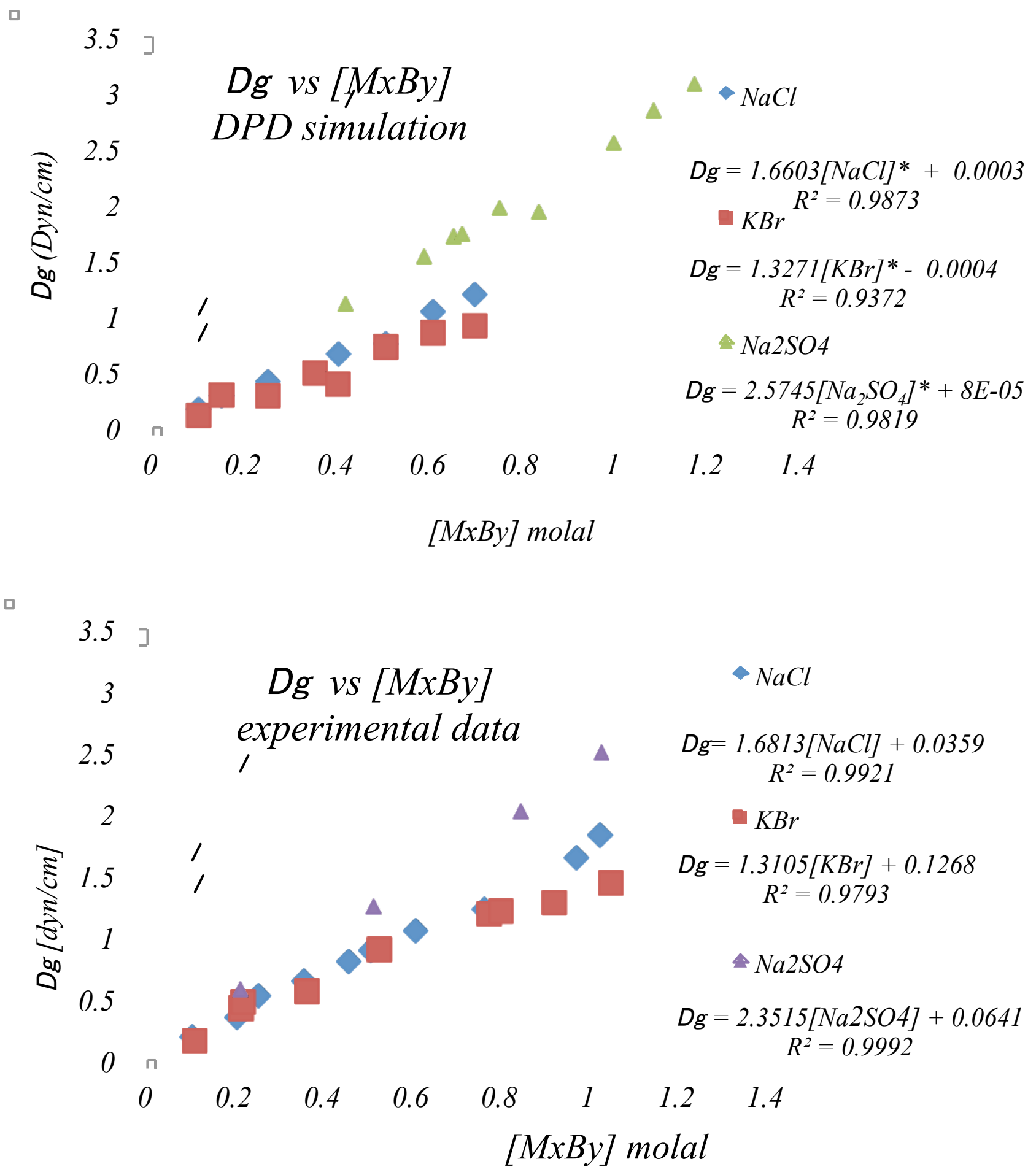

**Figure 4.** Increase of Interfacial tension between n-dodecane and water when MxBy is added . Top: DPD simulation; botton: Experimental data [2].

## V.II DPD Simulation for n-dodecane/water with HCl

We performed DPD electrostatic simulations for the system n-dodecane and water with *HCl* added also using the DPD electrostatic methodology described in Section IV. The repulsive parameters $a_{ij}$ for this system were obtained considering the dependence of Flory-Huggins $\chi$ parameters with the concentration via the activity coefficients and were calculated as had been discussed. The Flory-Huggins $\chi$ parameter for *HCl* and n-dodecane turns out to be:

$$\left[\sqrt{\chi_{HClw}} - \sqrt{\chi_{wo}}\right]^2 = \frac{v_m}{RT}(\delta_{HCl} - \delta_o)^2 \equiv \chi_{HClo}. \tag{31}$$

Three $H^+$ and $Cl^-$ ions were mapped in to a DPD bead with its respective charge distributed as in reference [16, 17].We added different quantities of DPD particles of *HCl*. Figure 5 shows the results obtained for the DPD simulation. As we can observe a very different behavior is obtained. In this case, the interfacial tension decreases as the concentration of *HCl* is increased. This behavior corresponds correctly with the experimental results reported for the addition of *HCl* in organic/water mixtures [4, 5].

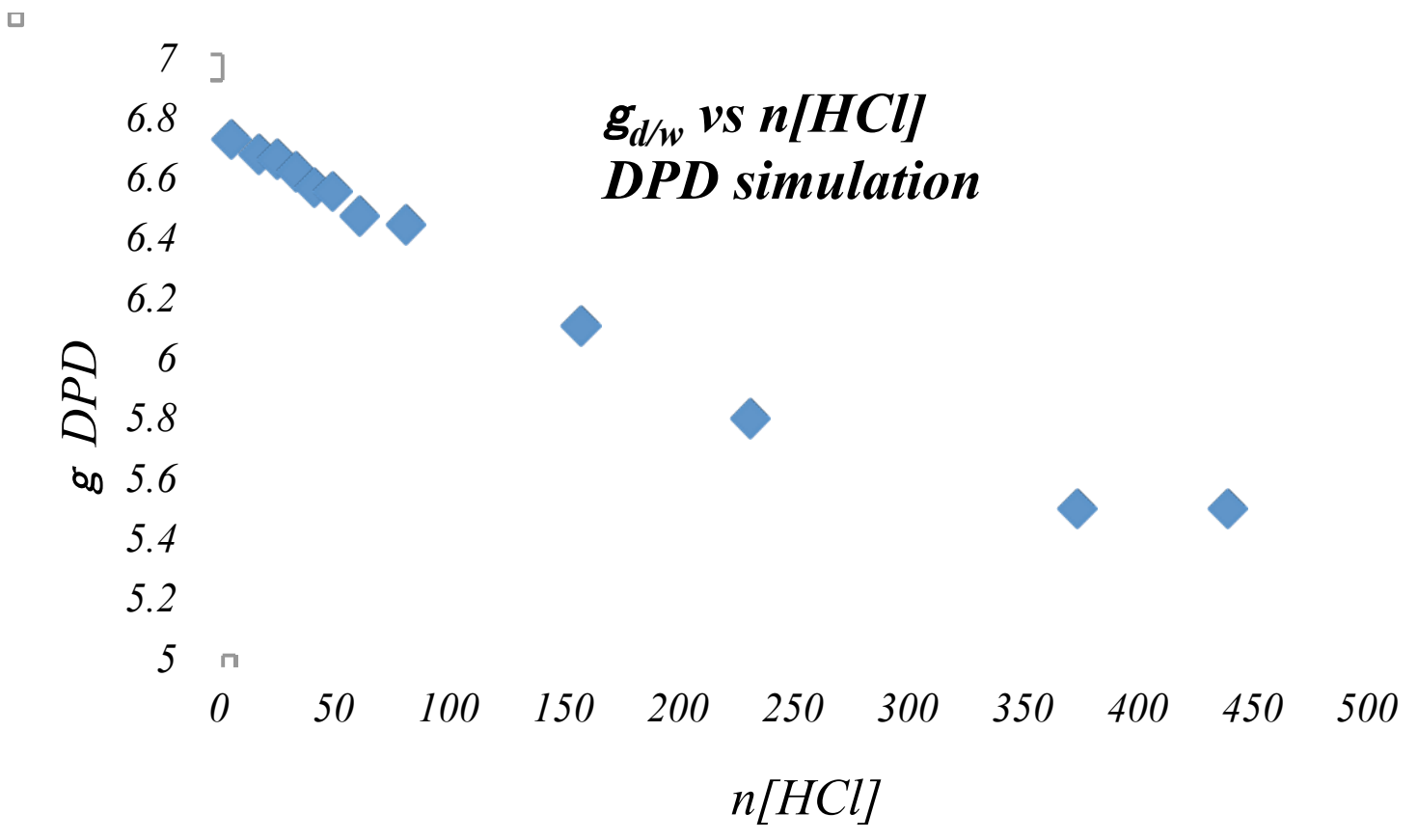

**Figure 5.** Interfacial tension for n-dodecane-water with *HCl* obtained by DPD simulations.

## VI CONCLUSIONS

The concentration dependence of the Flory-Huggins $\chi$ parameter suggests that the repulsive parameters $a_{ij}$ in the conservative force component used in DPD simulations must also depend on concentration. A correct parametrization for these quantities is then obtained through the activity coefficients. DPD simulations for obtaining the interfacial tension of an organic solvent in the presence of an electrolyte aqueous solution were carried out, following the methodology introduced by Gonzalez-Melchor and one of us [17], and using the concentration–dependent repulsive parameters previously calculated. Our simulations reproduce very well previously reported experimental results [2] for all the inorganic salts tried, observing an increase in the interfacial tension with salt concentration. Analyzing the density profiles, it would be clear that the ion distribution is heavily biased towards the bulk of the aqueous medium, leaving only a small amount at the interface, this is a consequence of this kind of ions being negatively adsorbed [2, 4, 5] as is shown in Figure 6. The ion interface depletion is normally explained by them being repelled away by image charges (cf. e.g. Onuki [7, 8]), amounting to a negative adsorption.

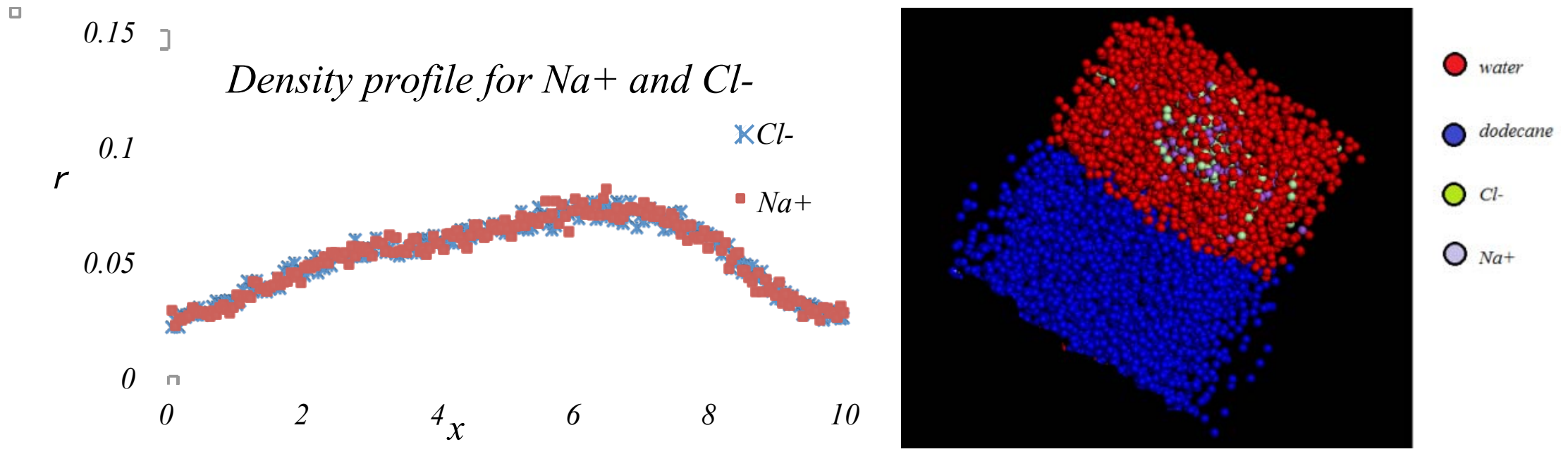

**Figure 6.** (Color on line) Density profile (left) and ion distribution (right) of $Na^+$ (violet) and $Cl^-$ (green) particles, shown to lie predominantly in the aqueous bulk (red particles).

Furthermore, when adding *HCl* to water in the presence of an organic solvent, the inverse behavior for the interfacial tension reported in the literature [4, 5] is also well reproduced by our results. This suggests that it is appropriate to consider concentration-dependent parameters in the conservative force component, as done here. It also shows that the use of activity coefficients to obtain the said parameters is a good alternative for the general class of systems studied.

Our parametrization was based on experimental data for the activity coefficients which involve a free energy dependence with concentration, and is therefore an adequate description of the interactions. This is the same that we would obtain from other theoretical models, such as that of Li and Lu [4, 5], which calculates the activity coefficients through the Meissner [6] method and which obtains the variation in the interfacial tension between an organic solvent and aqueous multielectrolyte solutios from the isothermal Gibbs equation and by using a Langmuir-type adsorption equation. We followed this same calculation and obtained essentially the same $a_{ij}$ parameters as the ones used in this work. Since we also included $UO_2Cl_2$ in our study, and since activity coefficients are much easier to find in the literature, by using experimental data to fix our parametrization we cover a much broader range of species.

As in polymeric mixtures the interfacial tension is very sensitive to variations in the electrolyte concentration, the present work should prove useful for the study of these systems. It is also of use for product design in industrial applications where oil/water interfaces are involved and the use of additives strong modifies the product characteristics. Finally the present study supports the use of the electrostatic DPD methodology.

## ACKNOWLEDGMENTS

We are grateful to A. Gama Goicochea and I. Soto Escalante for comments and discussions and to the DGTIC-UNAM for the additional computational support.

**BIBLIOGRAPHY**

[1] B. Cai, J. Yang, and T. Guo, J. Chem. Eng. Data **41**(3), 493 (1996); R.N. Healy, and R.L. Reed, Soc. Pet. Eng. J. **257,** 419, (1975).

[2] R. Aveyard and S.M. Saleem, J. Chem. Soc., Faraday Trans. **1** (72), 1609, (1976).

[3] C. Desnoyer, O. Masbernat, and C. Gourdon, J. Colloid Interface Sci. **191**, 22 (1997).

[4] Z. Li and B.C.-Y. Lu, Chem. Eng. Sci. **56**, 2879 (2001).

[5] Z. Li and B.C. –Y Lu, Fluid Phase Equilib. **200**, 239, (2002).

[6] H. P. Meissner and J.W. Tester, Ind. Eng. Chem. Proc. Des. Dev. **11**, 128, (1972).

[7] A. Onuki, Phys. Rev. E **73**, 021506 (2006).

[8] A. Onuki, J. Chem. Phys. **128**, 224704 (2008).

[9] M. Bier, J. Zwanikken, and R. van Roij, Phys. Rev. Lett. **101**, 046104 (2008).

[10] B. Rotenberg, I. Pagonabarraga, and D. Frenkel, Faraday Discuss. **144**, 223 (2010).

[11] P.J. Hoogerbrugge and J.M. Koelman, Europhys. Lett. **19**,155 (1992).

[12] R.D. Groot and P.B. Warren, J. Chem. Phys. **107**, 4423 (1997).

[13] K.P. Travis, M. Bankhead, K. Good, and S.L. Owens, J. Chem. Phys. **127,** 014109 (2007).

[14] J.H. Hildebrand and S.E. Wood, J. Chem. Phys. **1**, 817 (1933).

[15] G. Scatchard, Chem. Rev. **8**, 321 (1931).

[16] R.D. Groot, J. Chem. Phys. **118**, 11265 (2003).

[17] M. Gonzalez-Melchor, E. Mayoral, M.E. Velazquez, and J. Alejandre, J. Chem. Phys. **125**, 224107 (2006).

[18] A. Ghoufi and P. Malfreyt, J. Chem. Theory Comput. **8**, 787, (2012).

[19] A. Ghoufi and P. Malfreyt, Phys. Rev. E **83**, 051601, (2011).

[20] R.M. Pytkowicz, *Activity Coefficients in electrolyte solutions*" (CRC Press 1979).

[21] V.M. Lobo, *Handbook of Electrolyte Solutions* (Elsevier, Amsterdam 1989).

[22] M. Muthukumar, J. Hua, and A. Kundagrami, J. Chem. Phys. **132**, 084901 (2010).

[23] P. Español and P. Warren, Europhys. Lett. **30** (4), 191 (1995).

[24] D. Frenkel and B. Smith, *Understanding Molecular simulation: From Algorithms to Applications* (Academic Press, N.Y. 1996).

[25] I.V. Pivkin and G. E. Karniadakis, J. Chem. Phys. **124**, 184101 (2006).

[26] A. Maiti and S. McGrother, J. Chem. Phys. **120** (3), 1594 (2003).

[27] P.J. Flory, J. Chem. Phys. **10**, 51, (1942).

[28] M.L. Huggins, J. Chem. Phys. **9,** 440 (1941).

[29] M.G. Bawendi and K.F. Freed, J. Chem. Phys. **84** (12), 7036 (1986).

[30] P.B. Warren, J. Physique II **7**, 343, (1997).

[31] M.G. Bawendi, K.F. Freed, and U. Mohanty, J. Chem. Phys. **87** (9), 5534 (1987).

[32] J.H. Hildebrand and R.L. Scott, *Solubility of Nonelectrolytes* (Reinhold, New York, 1950).

[33] V.A. Baulin and A. Halperin, Macromolecules **35**, 6432 (2002).

[34] V.A. Baulin and A. Halperin, Macromol. Theory Simul. **12**, 549 (2003).

[35] A. Gama, M. Romero-Bastida, and R. López-Redón, Molec. Phys. **105** (17-18), 2375 (2007).

[36] R.F. Blanks and J.M. Prausnitz, I&EC Fundamentals, **3** ,1 (1964).

[37] A. Poisson and J. Chanu, Limnology and Oceanography, **21** (6) , 853 (1976).

[38] L.N. Trevani, E.C. Balodis, and P.R. Tremaine, J. Phys. Chem. B. **111**, 2015 (2007).

[39] A.F.M. Barton, Chem. Rev. **75** (6), 731 (1975).

[40] N.P. Allen and D.J. Tildesley, *Computer simulation of liquids* (Oxford University Press, New York, 1987).

[41] G.J. Gloor, G. Jackson, F.J. Blas, and E. de Miguel, J. Chem. Phys. **123**, 134703 (2005).

[42] One should be aware, however, of the fact that the use of pair-wise potentials without a proper reference to the coarse-graining procedure may lead to results which cannot be correctly interpreted (cf. e.g. A.A. Louis, J.Phys.: Condens. Matter **14**, 9187 (2002)).